\documentclass{aa}  

\usepackage{graphicx}
\usepackage{color}
\usepackage{txfonts}

\begin{document}

   \title{History and destiny of an emerging early-type galaxy}

   \subtitle{New IFU insights on the major-merger remnant NGC\,7252}

   \author{J. Weaver\inst{1,2}
          \and
          B. Husemann\inst{2}
          \and
          H. Kuntschner\inst{3}
          \and
          I. Martín-Navarro\inst{4,2}
          \and 
          F. Bournaud\inst{5}
	  \and
	  P.-A. Duc\inst{6,5,7}
	  \and
	  E.~Emsellem\inst{3,8}
	  \and
	  D.~Krajnovi\'c\inst{9}
	  \and
	  M.~Lyubenova\inst{3}
	  \and
	  R.~M.~McDermid\inst{10,11}
          }

	 \institute{SUPA, School of Physics and Astronomy, University of St Andrews, North Haugh, St Andrews KY16 9SS, UK\\ \email{jrw20@st-andrews.ac.uk}
         \and
         Max-Planck-Institut für Astronomie, K{\"o}nigstuhl 17, 69117 Heidelberg, Germany
         \and
         European Southern Observatory, Karl-Schwarzschild-Strasse~2,85748 Garching, Germany
         \and
         University of California Observatories, 1156 High Street, Santa Cruz, CA 95064, USA
         \and 
         Laboratoire AIM Paris-Saclay, CEA/IRFU/SAp, Universit\'e Paris Diderot, 91191 Gif-sur-Yvette Cedex, France
         \and 
         Universit\'e de Strasbourg, CNRS, Observatoire Astronomique de Strasbourg, UMR 7550, 67000 Strasbourg, France
         \and
         Universit\'e Paris Diderot, AIM, Sorbonne Paris Cit\'e, CEA, CNRS, 91191 Gif sur Yvette, France
         \and 
         Observatoire de Lyon, Centre de Recherche Astrophysique de Lyon and Ecole Normale Sup\'erieure de Lyon, Universit\'e Lyon 1, 9 Avenue Charles Andr\'e, 69230 Saint-Genis Laval, France
         \and 
         Leibniz--Institut f{\"u}r Astrophysik Potsdam (AIP), An der Sternwarte 16, 14482 Potsdam, Germany
         \and 
         Department of Physics and Astronomy, Macquarie University, Sydney, NSW 2109, Australia
         \and
         Australian Astronomical Observatory, PO Box 915, Sydney, NSW 1670, Australia
    }
   \date{}

 
  \abstract
   {The merging of galaxies is one key aspect in our favourite hierarchical $\Lambda$CDM Universe and is an important channel leading to massive quiescent elliptical galaxies. Understanding this complex transformational process is ongoing.}
   {We aim to study NGC\,7252, which is one of the nearest major-merger galaxy remnants, observed $\sim$1\,Gyr after the collision of presumably two gas-rich disc galaxies. It is therefore an ideal laboratory to study the processes inherent to the transformation of disc galaxies to ellipticals.}
   {We obtained wide-field IFU spectroscopy with the VLT-VIMOS integral-field spectrograph covering the central $50\arcsec \times 50\arcsec$ of NGC\,7252 to map the stellar and ionised gas kinematics, and the distribution
and conditions of the ionised gas, revealing the extent of ongoing star formation and recent star
formation history.}
   {Contrary to previous studies, we find the inner gas disc not to be counter-rotating with respect to the stars. In addition,  the stellar kinematics appear complex with a clear indication of a prolate-like rotation component which suggests a polar merger configuration. The ongoing star formation rate is $2.2\pm0.6$ $M_{\odot}\,\mathrm{yr}^{-1}$ and implies a typical depletion time of $\sim$2\,Gyr given the molecular gas content. Furthermore, the spatially resolved star formation history suggests a slight radial dependence, moving outwards at later times. We confirm a large AGN-ionised gas cloud previously discovered $\sim$5 kpc south of the nucleus, and find a higher ionisation state of the ionised gas at the galaxy centre relative to the surrounding gas disc. Although the higher ionisation towards the centre is potentially degenerate within the central star forming ring, it may be associated with a low-luminosity AGN.}
 {Although NGC\,7252 has been classified as post-starburst galaxy at the centre, the elliptical-like major-merger remnant still appears very active. A central kpc-scale gas disc has presumably re-formed quickly within the last 100\,Myr after final coalescence. The disc features ongoing star formation, implying Gyr long timescale to reach the red sequence through gas consumption alone. While NGC~7252 is useful to probe the transformation from discs to ellipticals, it is not well-suited to study the transformation from blue to red at this point.}


   \maketitle
%

\section{Introduction} \label{Introduction}

Mergers of galaxies are a natural consequence of the hierarchical build-up of large-scale structure, and emergent from our concordance $\Lambda$ cold dark matter ($\Lambda$CDM) cosmological model. This model has been shown to provide  a good working paradigm to describe the formation and evolution of dark matter haloes and their galaxies within \citep[e.g.][]{Springel:2005b}. In particular, the merger of two late-type (disc) galaxies has become a major scenario to assemble massive early-type (elliptical) galaxies, as shown by early numerical simulations \citep[e.g.][]{Hernquist:1991,Barnes:1992}. Despite the success of this scenario, we have yet to fully understand the various processes at play during major mergers, made complex by intricate baryonic physics, radiative feedback, and the hydrodynamics  and chemistry of the gas.

Since the pioneering work to understand the role of hierarchical merging in the formation of early-type galaxies, numerical simulations have continued to mature due to increased resolution, implementation of more refined and extended prescriptions of baryonic physics, and larger statistical samples \citep[e.g.][]{Naab:2003, Bournaud:2005, diMatteo:2008, Genel:2014, Vogelsberger:2014, Vogelsberger:2014b,Schaye:2015,Tsatsi:2017, Li:2018}. These and other inclusions now allow for quantitative predictions of various merger remnant properties. One major issue identified by the simulations is that massive early-type galaxies appear too blue, due to high continuous star formation resulting from gas inflow during the merger \citep[e.g.][]{Hopkins:2013}. This has often been addressed by invoking energetic feedback from active galactic nuclei (AGN) which too are ignited following gas inflow, but subsequently expel gas from their host galaxy core and hence suppress star formation \citep[e.g.][]{Matteo:2005, Springel:2005, Somerville:2008}. This has lead to a popular scenario in which AGN play a key role in shaping the properties of early-type galaxies during their assembly, following from a two disc major merger \citep[e.g.][]{Sanders:1988b, Hopkins:2008a, Schawinski:2010b}.

Observationally, integral-field spectroscopy has provided a significant step forwards in understanding early-type galaxy formation by recovering the 2D projected internal dynamics and properties, and hence a fossil record of their formation process. Integral-field unit (IFU) surveys covering a large number of early-type galaxies, such as SAURON \citep{Zeeuw:2002}, ATLAS${}^\mathrm{3D}$ \citep{Cappellari:2011}, CALIFA \citep{Sanchez:2012a}, SAMI \citep{Croom:2012}, MaNGA \citep{Bundy:2015}, or the MASSIVE surveys \citep{Ma:2014}, have revealed a rich variety of kinematic properties and sub-structures  \citep[e.g.][]{Emsellem:2004,Emsellem:2011,Krajnovic:2008,Krajnovic:2013,Falcon-Barroso:2017, Veale:2017,vandeSande:2017}. Based on the 2D mapped kinematics, the specific angular momentum has been proposed as a metric for classifying fast and slow rotating early-type galaxies \citep{Cappellari:2007,Emsellem:2011}. While dedicated numerical simulations have shown that this classification does not reflect a bimodality in their merger histories \citep{Naab:2014}, differences between these populations remain. Observationally, the fraction of fast to slow rotators seems to be a strong function of stellar mass \citep{Emsellem:2011,Veale:2017} and fast rotators frequently show strong evidence of disc-like structures, often with increased metallicity and younger ages \citep{Krajnovic:2008,Kuntschner:2010}. These populations also have been identified as having different star formation histories and star formation quenching timescales \citep{Smethurst:2018}.

While the transformation from discs to an elliptical galaxy can be readily addressed through simulations guided mainly by gravitation, the star formation histories are related to complex baryonic physics which are difficult to fully implement. Indeed, early-type galaxies are known to show a great variety of ionised \citep[e.g.][]{Sarzi:2006,Singh:2013}, atomic \citep[e.g.][]{Serra:2012} and molecular gas content \citep[e.g.][]{Young:2011}, which can have an internal or external origin, as discussed in \citet{Davis:2011}, which depend on environment and merger history. This highlights the complexity in the star formation histories and associated processes responsible for transforming blue (star-forming) disc galaxies into red (quiescent) elliptical galaxies. While fast quenching of early-type galaxies has been often proposed in light of AGN feedback \citep[e.g.][]{Schawinski:2010b}, inconsistent quenching timescales of several Gyr have also been observed \citep[e.g.][]{Weigel:2017,Smethurst:2018}.

It is therefore still necessary to observe and characterise ongoing major mergers at various evolutionary phases in order to establish a complete picture of their transformation. Given the decreasing major merger rate towards lower redshift  \citep[e.g.][]{Lotz:2011,Rodriguez-Gomez:2015}, these events are particularly rare in the local Universe. This scarcity has driven detailed studies of spatially resolved nearby major mergers such as the Mice \citep[NGC~4676; e.g.][]{Barnes:2004,Wild:2014}, the Antennae \citep[NGC~4038,NGC~4039; e.g.][]{Whitmore:2010,Ueda:2012}, or the Atoms-for-peace galaxy \citep[NGC~7252; e.g.][]{Schweizer:1982, Wang:1992, Hibbard:1994, Schweizer:2013}. While these individual galaxies are not necessarily representative of the entire population and the total parameter space concerning major mergers, they nonetheless provide important information about the transformational processes. 

In this article, we focus on \object{NGC\,7252} ($z=0.016$) aiming to uncover the history and destiny of this enigmatic merger remnant based on new wide-field optical IFU observations. The system is well studied with many ancillary observations available, providing a well-defined framework from which to interpret our new observations. Bright tidal tails are clearly visible around NGC\,7252, which are rich in atomic \ion{H}{i} gas \citep{Hibbard:1994}. Numerical simulations suggest that these features are produced by a major merger of two gas-rich galaxies which rarely last longer than $\sim$\,500\,Myr \citep[e.g.][]{Duc:2013} after the first passage of the galaxies. A dedicated numerical simulation of NGC\,7252 has been presented by \citet{Chien:2010} to study the star formation history during this merger. Their simulation predicts a rise in the star formation rate (SFR) when the two galaxies approached the first close passage at pericentre about $\sim$\,620\,Myr ago. Another burst in star formation is associated with the second encounter about $\sim260$\,Myr ago, followed by the final coalescence of the nuclei about $215$\,Myr ago. A prolonged star-formation episode lasting for $\sim60$\,Myr is predicted by the simulations without significant levels of star formation thereafter. While the stellar population at the very central galaxy core is indeed characterised by a post-starburst spectrum \citep{FritzeAlvensleben:1994}, it is surrounded by a rotating disc of molecular \citep{Wang:1992,Ueda:2014} and ionised gas \citep{Schweizer:1982} within $\sim$\,8" (2.4 kpc) which is still actively forming stars. 
The surface brightness profile of NGC~7252 has already become well-described by a r$^{1/4}$ law \citep[e.g.][]{Schweizer:1982, Rothberg:2004}, suggesting that NGC~7252 is close to finishing the transformational process from two disc galaxies to an elliptical galaxy.

The paper is organised as follows: in section~\ref{Observations} we give a brief report of the observations and data reduction. This is followed by the section~\ref{Analysis_and_Results} where we present the analysis and results which are discussed in more detail in section~\ref{Discussion}. The paper finishes with a summary and conclusion in section~\ref{Conclusions}. Throughout the paper we adopt a concordance cosmological model with $H_0=70\,\mathrm{km}\,\mathrm{s}^{-1}\mathrm{Mpc}^{-1}$, $\Omega_\mathrm{m}=0.3$, and $\Omega_\Lambda=0.7$.

\begin{figure}
\centering
\includegraphics[width=\hsize]{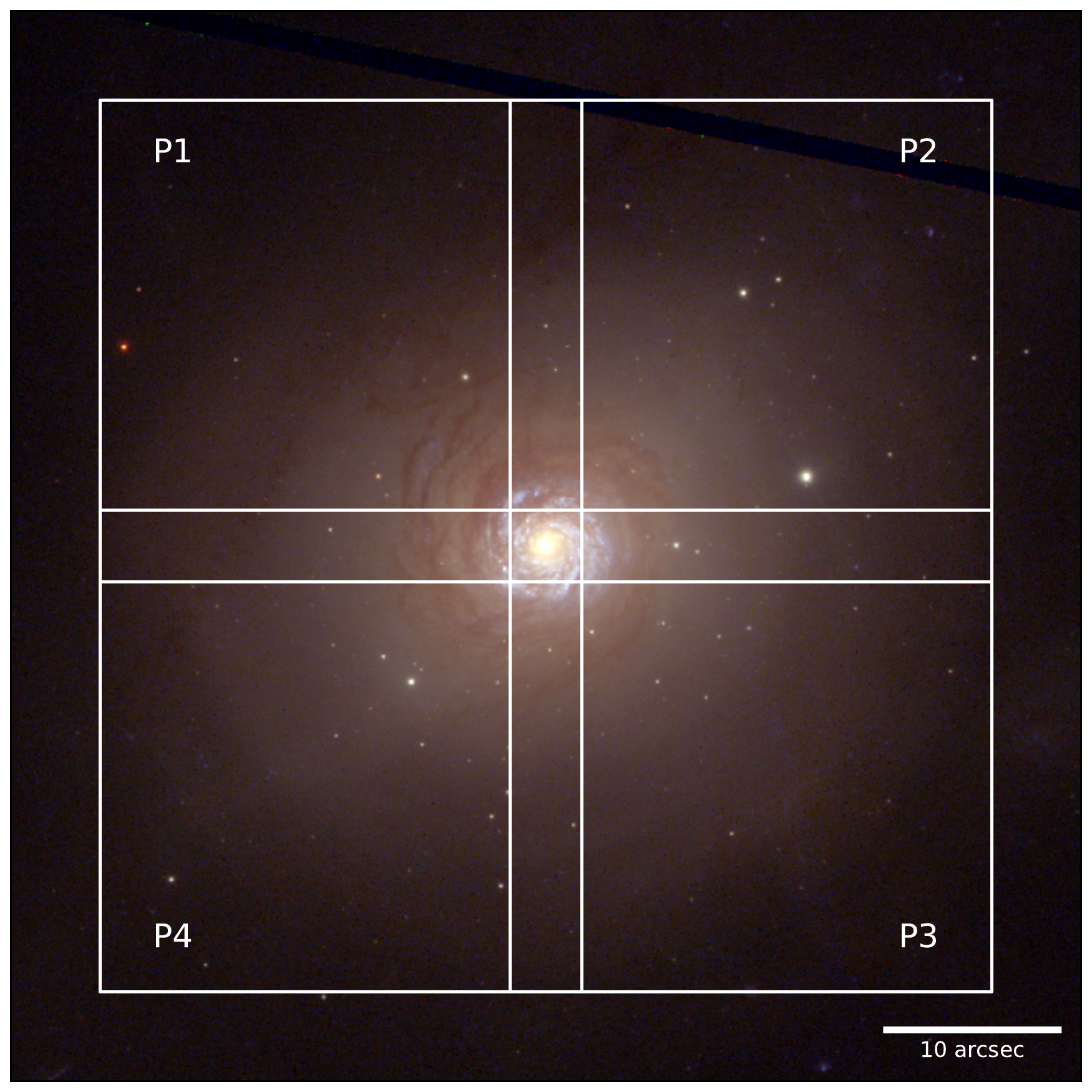}
\caption{Field-of-view of the four independent VIMOS pointings. A white box denotes each pointing, overlaid on a colour image of NGC\,7252 taken with WFC3 aboard \textit{Hubble} in the bands F336W, F475W and F775W \citep{Bastian:2013}.}
\label{fig:VIMOS_pointings}
\end{figure}

\section{Observations and data reduction} \label{Observations}

We observe the central $50\arcsec \times 50\arcsec$  of NGC\,7252 with the VIsible MultiObject Spectrograph (VIMOS) instrument \citep{LeFevre:2003} using the integral-field unit (IFU) mode in October 2011 as part of programme 088.B-0224 (PI: H. Kuntschner). With the high-resolution blue grating of VIMOS we cover a wavelength range from 4130\AA\ to 6200\AA\ with a spectral resolution of $R\sim2550$ over a $27\arcsec\times27\arcsec$ field-of-view per individual pointing. Hence, we cover the main body of NGC\,7252 with four pointings that slightly overlap to always cover the centre of the galaxy as shown in Fig.~\ref{fig:VIMOS_pointings}. Each pointing is observed twice for 1500\,s together with 500\,s long blank sky field exposure. Three lamp flat exposures as well as one arc lamp exposure are observed as night time calibrations. Standard star observations for our setup are taken as part of the standard calibration plan.

The data reduction is performed entirely with the \textsc{Py3D} data reduction package \citep{Husemann:2013b} that has been developed for fibre-fed IFUs as part of the Calar Alto Large Integral Field Area survey \citep[CALIFA,][]{Sanchez:2012a}. It has already been successfully used to reduce several VIMOS IFU data sets \citep{Husemann:2014,Woo:2014,Richtler:2017} and we follow the same scheme here, as briefly described below. 

After bias subtraction, cosmic rays are detected with \textsc{PyCosmic} \citep{Husemann:2013b} and flagged throughout the processing. Fibres are identified in the combined lamp images and the cross-dispersion width of fibre profiles are estimated as a function of wavelength to allow for an optimal extraction \citep{Horne:1986} of the fibre counts, even in the presence of severe cross-talk between fibres. From the arc lamp observations we determine the wavelength solution for each fibre as well as the spectral resolution which can vary significantly across the four independent spectrographs. We applied the wavelength solution to the data after extraction and adaptively smooth the data to a common spectral resolution of 3\,\AA, corresponding to the worst resolution found among the data sets. Fibre-to-fibre transmission differences are corrected using a fibre-flat created from the continuum lamp observations of each corresponding observing block. Flexure of the instrument is handled by measuring the shifts of sky line spots in the science data cross-dispersion and dispersion direction. Flux calibration is performed by reducing standard star observations in exactly the same way as the data from which a response function is determined by comparing the observed counts with the reference spectrum. 

\begin{figure*}
      \resizebox{\hsize}{!}{\includegraphics{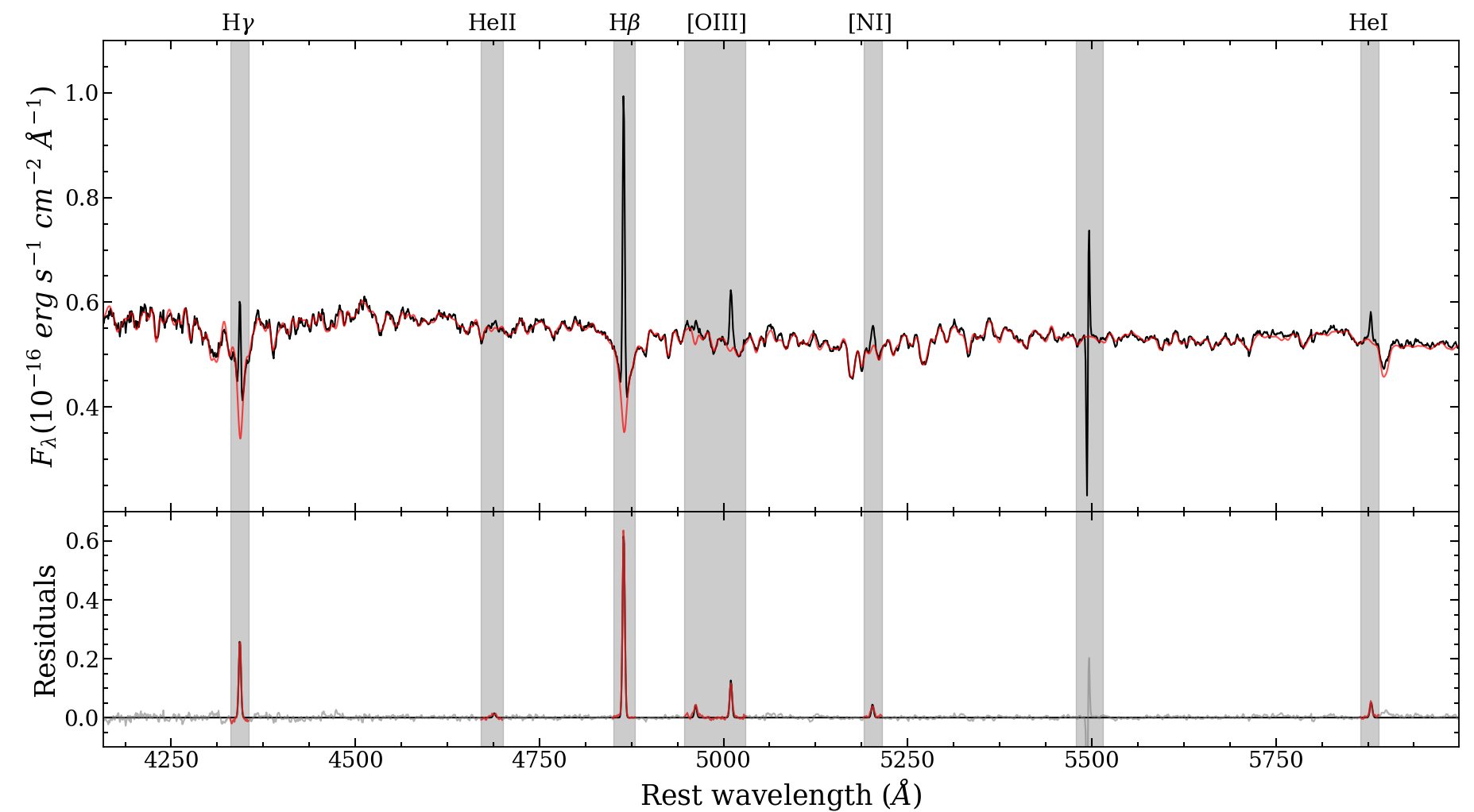}}
      \caption{Example spectral modelling of the brightest spaxel at the galaxy centre. The top panel shows the stellar continuum (black), fitted with a model spectrum (red). Regions around emission lines (grey) are omitted from the continuum fit. The bright \ion{O}{i} sky line at 5577\,\AA\ (observed frame) is also omitted. The bottom panel shows the emission line model (red), obtained by fitting the residuals (grey) of the spaxel and continuum model. Rest-frame transformation is calculated from cosmological redshift alone ($z=0.016$).}
      
      \label{fig:spectum_fitting}
   \end{figure*}
   
In post-processing we subtract the sky background as measured from the reduced blank sky field closest in time. Due to varying levels of stray light, the background is subtracted independently in each spectrograph quadrant. The effect of differential atmospheric refraction is accounted for by tracing the bright centre of the galaxy as a function of wavelength in each observation. The final datacube measuring $50\arcsec\times50\arcsec$ is then re-constructed from the individual flux-calibrated and sky-subtracted pointings using the \textsc{drizzle} algorithm \citep{Fruchter:2002}. The galaxy centre serves as the reference point to align all pointings over the entire wavelength range.

When computing absolute quantities from the resulting spectra in the IFU data, a good absolute spectrophotometric calibration is key. We check the accuracy of our absolute spectrophotometric accuracy by comparing the IFU data with the photometry of the available {\it Hubble} images in the F475W filter, corresponding to Sloan $g$ band. Our VIMOS spectral range covers more than 90\% of the effective F475W filter band. We measure $m_\mathrm{IFU}= 13.0$\,mag (AB) compared to $m_\mathrm{HST}=12.9$\,mag (AB) within an aperture of 30\arcsec\ in diameter centred on the nuclear region of NGC~7572. Hence, we adopt a systematic photometric error of $\leq10\%$ for all absolute quantities inferred from the VIMOS data.

\begin{figure*}
      \resizebox{\hsize}{!}{\includegraphics{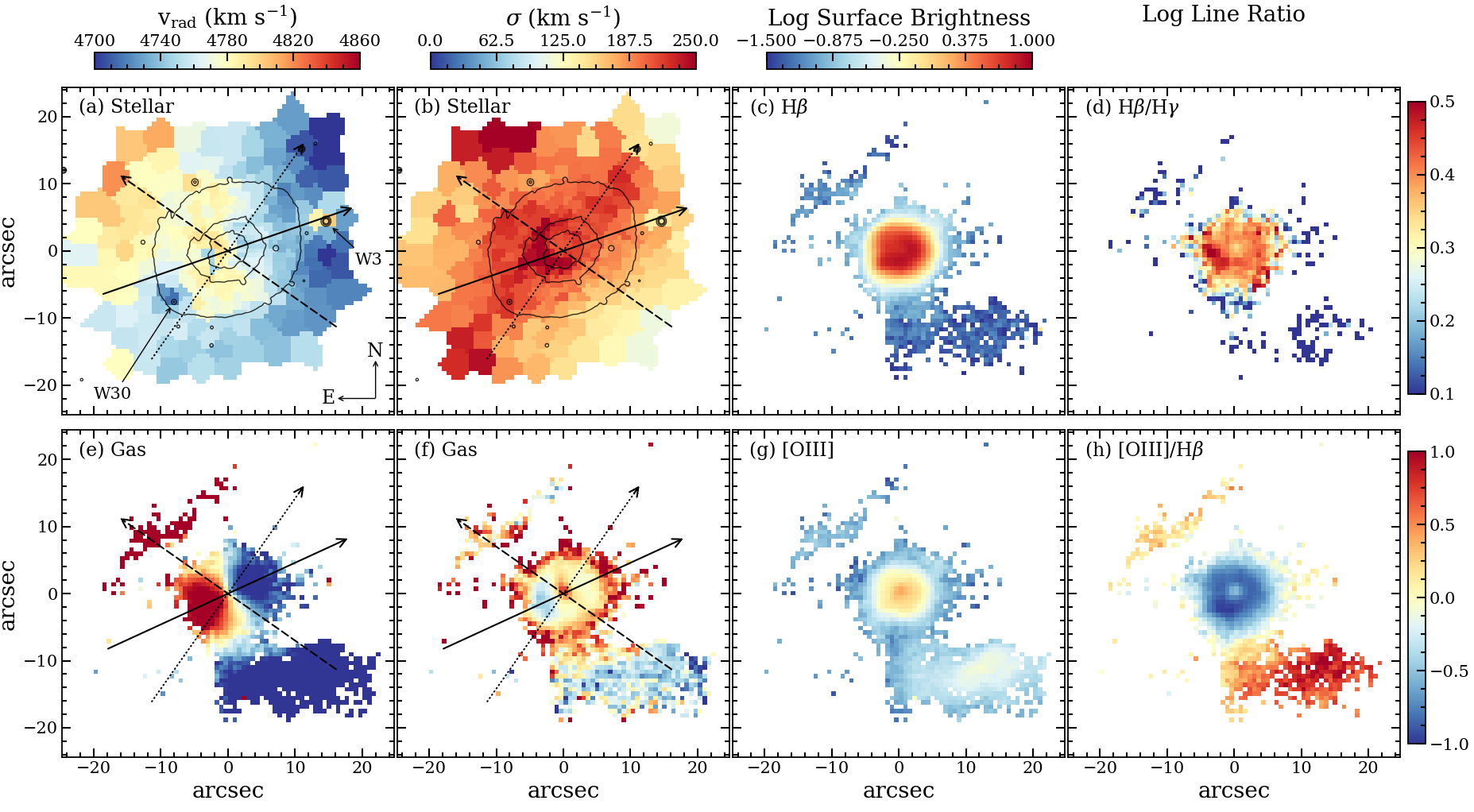}}
      \caption{Kinematic and emission line overview. The stellar maps in \textit{panels} \textit{a} and \textit{b} are derived from kinematic measurements on 135 Voronoi co-added spectra using \textsc{PyParadise}. Overlaid black contours at 24, 25, and 26 $\mathrm{mag\,arcsec}^{-2}$ (AB) are constructed from smoothed F775W \textit{Hubble} imaging \citep{Bastian:2013}. The gas kinematic maps in \textit{panels} \textit{e} and \textit{f} are derived from emission line measurements on each spaxel, excluding pixels with low S/N or high velocity and dispersion errors. Position angles are indicated by arrows. The major axis of gas and stellar kinematics agree within 6\degr, shown both as solid lines. The axis of the prolate-like rotation is shown as the dashed line and the photometric PA is shown by the dotted line as measured from  elliptical isophotes at $>$15\arcsec. The gas flux maps are shown in \textit{panels c} and \textit{g}, in units of $\mathrm{Log(}10^{-16}\mathrm{erg\,s}^{-1}\mathrm{cm}^{-2}\mathrm{)}$. The Balmer ratio map in \textit{panel d} serves as an approximate extinction map and the variation in ionisation hardness is highlighted by the [OIII]/H$\beta$ ratio in \textit{panel g}.}
      \label{fig:kinematic_maps}
   \end{figure*}

\section{Analysis and results} \label{Analysis_and_Results}
\subsection{Stellar and emission line modelling} \label{Modelling}
The signal-to-noise ratio (S/N) of the stellar continuum varies significantly over the VIMOS field-of-view (FoV). Thus, in order to obtain the best stellar kinematics for the entire field, spectra are co-added using the Voronoi binning algorithm \citep{Cappellari:2003} with a target $S/N$ $\sim$\,50. This produces a total of 135 co-added spectra, with the largest bins near the edges of the field where the brightness is significantly less than at the field centre, as expected. 

Continua of the 135 spectra are then fitted with \textsc{PyParadise}, an extended Python version of \textsc{Paradise} \citep{Walcher:2015}. We adopt the input library of single stellar population spectra from the Medium-Resolution Isaac Newton Telescope Library of Empirical Spectra \citep[\textsc{MILES};][]{Sanchez-Blazquez:2006,Vazdekis:2010,Falcon-Barroso:2011}. The upper panel of Fig.~\ref{fig:spectum_fitting} is presented as an example, whereby the spectrum of the most central and brightest spaxel, which also corresponds to a single Voronoi bin, is fitted with the continuum model.

Emission line regions are masked during the continuum fitting and normalisation, highlighted in grey. The width of these exclusion regions allows for an emission line to remain masked despite the peculiar velocities of each bin. One may note the \ion{O}{i} bright sky line residual around 5577\,\AA\ (observed frame), which is also excluded. Although the spectra of most Voronoi bins agree with the target continuum S/N, the spectra of the edge regions, even though heavily binned, still exhibit unacceptably low S/N which is related to limited area per bin at the edges of the VIMOS FoV. These bins are discarded for clarity. The resulting stellar radial velocity and velocity dispersion maps are shown in panels a and b of Fig.~\ref{fig:kinematic_maps}, respectively.

Emission lines are extracted with \textsc{PyParadise} after the best-fit stellar continuum model is subtracted from the initial data. Given the higher S/N in the emission lines, we repeat the stellar continuum fitting on the unbinned cube but use the previously inferred Voronoi-binned stellar kinematics field as a fixed prior. Hence, only the linear superposition of stellar-population synthesis spectral basis is performed per spaxel and the best-fitting continuum model, at fixed stellar kinematics, is subtracted after which the emission lines are fitted in the residual spectrum. An example of the emission-line modelling for this two-step fitting process is shown in the lower panel of Fig.~\ref{fig:spectum_fitting} for the innermost spaxel, which also represents a single Voronoi bin given its high S/N. The emission lines are selected a priori in the rest frame and treated as Gaussian profiles. Kinematic error estimates are derived from a 30 trial bootstrapping on each spectrum. The resulting gas radial velocity and velocity dispersion maps are shown in panels e and f of Fig.~\ref{fig:kinematic_maps}. We also obtain line fluxes with \textsc{PyParadise}. The H$\beta$ and [\ion{O}{iii}] $\lambda 5007$ (shortly [\ion{O}{iii}] hereafter) surface brightness maps are shown in panels c and g, respectively. 

As with the continua fitting, the central regions produce high S/N line detections which rapidly deteriorate towards the edges of the field. Spaxels with S/N $<$~3, a radial velocity error $>$~60\,$\mathrm{km\,s}^{-1}$, or FWHM error $>$~150\,$\mathrm{km\,s}^{-1}$ are deemed insignificant and discarded. The pixel retention for the gas kinematic maps in panels e and f is carried over from the [\ion{O}{iii}] flux map in panel g.

\subsection{Ionised gas kinematic modelling}\label{gas_kinematics}
\begin{figure}
   \centering
   \includegraphics[width=\hsize]{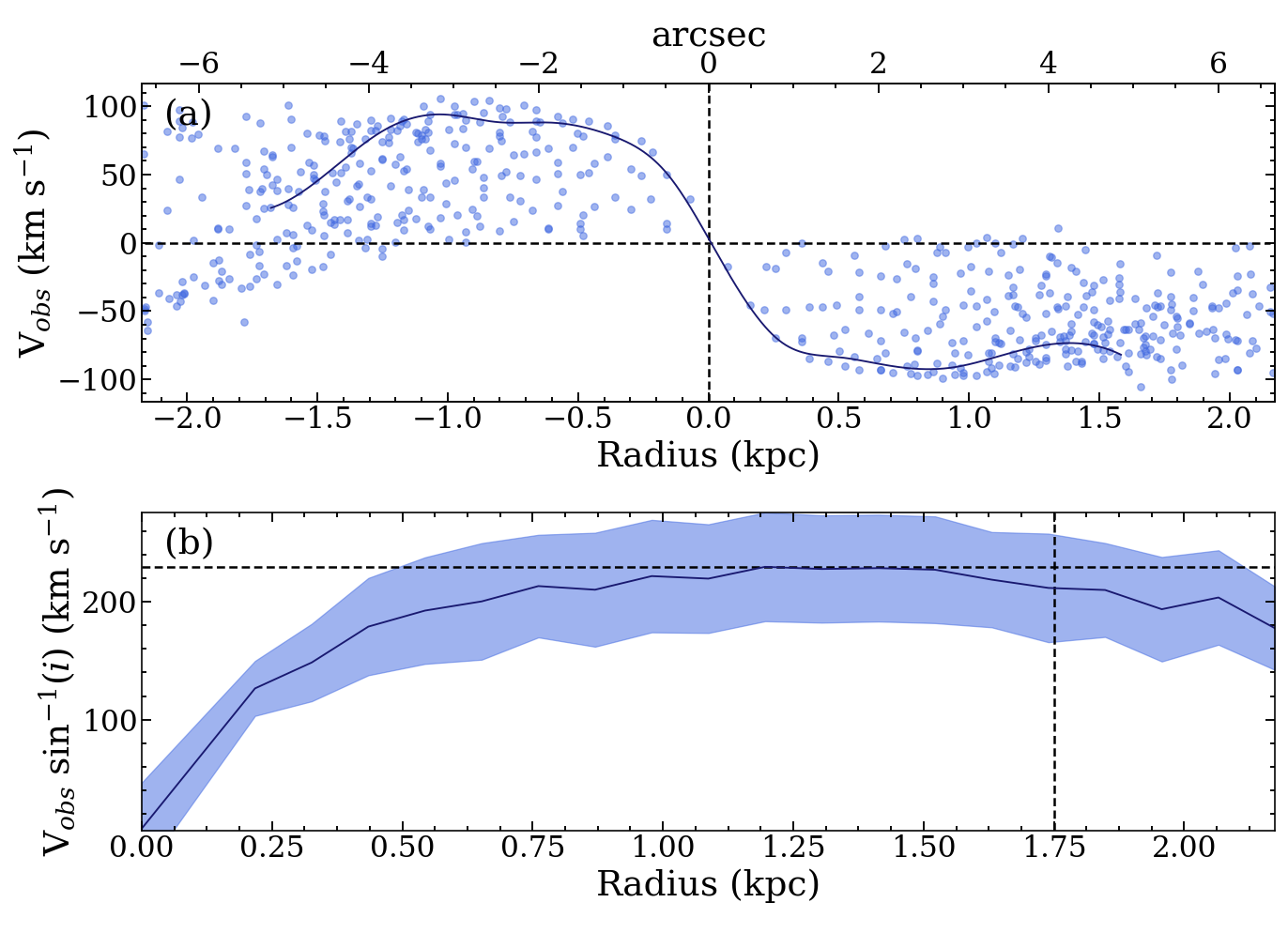}
      \caption{Velocity pattern of the nuclear star-forming disc. \textit{Panel a}: shows the $p$--$v$ diagram of the ionised gas kinematics of all pixels within the central 2\,kpc. It shows a nice rotational pattern with zero velocity along the minor axis. The velocity profile along the major axis from the \textsc{DiskFit} model is shown as the solid line. \textit{Panel b}: shows the inclination-corrected velocity curve derived from aperture fittings using \textsc{DiskFit}. The curve flattens around 1.25 kpc from the centre of the disc, at a circular velocity of $v_\mathrm{circ}=229\pm45 \mathrm{km\,s}^{-1}$. We attribute the fall off at large radii to noise. The intersecting dotted black lines indicate the point from which the dynamical mass of the disc is calculated, as discussed in the text.}
              
         \label{fig:velocity_curve}
   \end{figure}

One of the most interesting features of NGC\,7252 is a central rotating gas disc (see Fig.~\ref{fig:kinematic_maps} panels e and f) which exhibits a low velocity dispersion. The $p$--$v$ diagram (Fig.~\ref{fig:velocity_curve} upper panel) is typical for an inclined rotating disc and we examine the disc kinematics by fitting the radial velocity map using \textsc{DiskFit} \citep{Spekkens:2007,Sellwood:2015}. From \textsc{DiskFit}, we obtain a disc kinematic position angle $\mathrm{PA}_\mathrm{gas}=-65\degr\pm1\degr$ (counter-clockwise with respect to north, solid arrow in Fig.~\ref{fig:kinematic_maps} panels (e) and (f)), an ellipticity of 0.10$\pm$0.04, and a disc inclination of 26\degr$\pm$5\degr. Our disc inclination value agrees well with the most recent findings by \citet{Ueda:2014}, whereby the molecular gas disc is shown to have an inclination angle of 23\degr$\pm$3\degr. However, earlier results by \citet{Schweizer:1982}, obtained with slit spectra, find a conflicting inclination angle of 41\degr$\pm$9\degr.

We utilise the inclination-corrected aperture velocity measurements from \textsc{DiskFit} to investigate the velocity curve of the inner 6\arcsec\ (2 kpc) of the disc. As shown in Fig.~\ref{fig:velocity_curve}, the velocity curve flattens around 1.25 kpc to an inclination-corrected circular velocity of $v_\mathrm{circ}=229\pm45 \mathrm{km\,s}^{-1}$. The apparent decline of velocity in the outer annuli is significantly affected by higher noise and beam smearing effects at the edge of the visible gas disc. Whether the velocity curve would remain flat or really declines is therefore not possible to verify. The velocity of the inner gas disc leads to an enclosed dynamical mass of $M_\mathrm{dyn}(R<1.75\mathrm{kpc})=(2.1\pm0.9)\times10^{10}\,M_\sun$.

In addition to the inner gas disc, we detect outer gas streams to the north-east (NE) and south-west (SW) (Fig.~\ref{fig:kinematic_maps} panel e) which also appear to have a low velocity dispersion. As seen in the narrow-band imaging of \citet{Schweizer:2013}, these gas streams may be related to the larger gas tails which encircle the nucleus. The outer gas streams appear to have a velocity gradient nearly perpendicular to that of the inner gas disc, which initially led to the conclusion that the inner gas disc is counter-rotating with respect to the main body of the galaxy \citep[e.g.][]{Wang:1992}. However, if these gas streams are indeed related to the tidal tails this conclusion is not necessarily valid, as discussed further in section~\ref{stellar_kin}.

The significantly increased velocity dispersion of the gas disc towards the galaxy centre may be either related to shock-like ionisation as indicated by emission line diagnostics (see section~\ref{Ionisation_Conditions}), or simply caused by the beam smearing of the strong gas velocity gradient at the centre. The connection between gas excitation and velocity dispersion as commonly observed in the diffuse ionised gas phase in and around galaxies \citep[e.g.][]{Monreal-Ibero:2010, Ho:2014} is consistent with the first scenario. Indeed, it could be related to weak activity of an AGN as discussed in section~\ref{AGN_or_LINER}. On the other hand, high central gas dispersions are also a common feature in rotating gas discs due to the observational limitation of the beam smearing. Quantifying if beam smearing alone can explain the peak in velocity dispersion with sufficient precision is not possible in this case as we lack information on the intrinsic unsmoothed line flux distribution and a model of the PSF for these observations. In reality, we expect that both effect, beam smearing and change in physical condition, contribute to the high line dispersion at the nucleus.

\subsection{Stellar kinematics}\label{stellar_kin}
The stellar radial velocity map shows a clear east-west velocity gradient with a similar position angle and rotation as the circum-nuclear gas disc. This clearly suggests that the gas disc is not counter-rotating and indeed shares the same angular momentum vector as the stellar component. This rules out the gas counter-rotating with the stars as suggested by \citet{Wang:1992}. 

Similar to the methods described in section~\ref{gas_kinematics} for the gas disc, we use \textsc{DiskFit} to obtain the position angle of the stellar radial velocity component $\mathrm{PA}_\mathrm{stellar}$, measured over the same inner 6\arcsec\ (2 kpc). This measurement yields $\mathrm{PA}_\mathrm{stellar}=-71\degr\pm10\degr$ (counter-clockwise with respect to north, solid arrow in Fig.~\ref{fig:kinematic_maps} panels (a) and (b)). $\mathrm{PA}_\mathrm{stellar}$ is readily consistent with $\mathrm{PA}_\mathrm{gas}$. We find that the inclination as derived from the stellar radial velocity map ($i=18\pm23\degr$) are consistent with results from the gas disc ($i=26\pm5\degr$), but suffer from slightly larger error bars.

\begin{figure}
   \centering
   \includegraphics[width=\hsize]{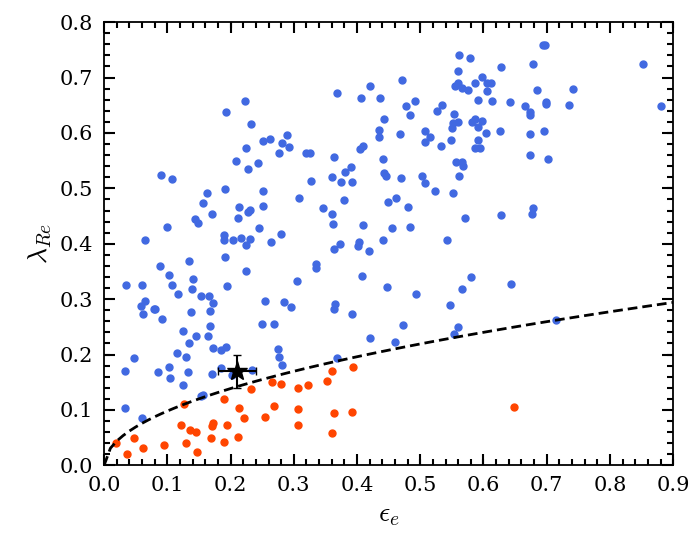}
      \caption{Specific angular momentum $\lambda_{R_\mathrm{eff}}$ against ellipticity $\epsilon_\mathrm{eff}$. The measured value for NGC~7252 $\lambda_{R_\mathrm{eff}}=0.17\pm0.03$ is shown as the black star and compared to the fast (blue) and slow (red) rotating early-type galaxies as obtained from the ATLAS$^\mathrm{3D}$ survey \citep{Emsellem:2011}. The black dashed line represents the proposed dividing line $\lambda_{R_\mathrm{eff}}=0.31\sqrt{\epsilon_\mathrm{eff}}$ proposed by \citet{Emsellem:2011}.}
         \label{fig:ATLAS3D}
   \end{figure}

To check whether NGC~7252 can be classified as a fast or slow rotator \citep[e.g.][]{Emsellem:2007,Emsellem:2011}, we computed the projected specific angular momentum $\lambda_R$ as described in \citep{Emsellem:2007} from the stellar velocity and velocity dispersion maps. Ancillary information on the effective radius ($R_\mathrm{eff}$), ellipticity ($\epsilon$) and photometric position angle ($\mathrm{PA}_\mathrm{phot}$) was obtained through ellipse isophotal fitting of the archival \textit{Hubble} F775W and F475W images using the Python Photutils package \citep{Bradley:2017}. Those measurements lead to $R_\mathrm{eff}=4.9\pm0.3$\,kpc, $\epsilon_\mathrm{eff}=0.21\pm0.03$ and $\lambda_{R_\mathrm{eff}}=0.17\pm0.3$ as well as $\mathrm{PA}_\mathrm{phot}=-35\degr$ (counter-clockwise with respect to north, dotted arrow in Fig.~\ref{fig:kinematic_maps}). In Fig.~\ref{fig:ATLAS3D}, we compare our measurements with those of early-type galaxies from $\mathrm{ATLAS}^\mathrm{3D}$ as published by \citet{Emsellem:2011}. We find that NGC~7252 lies very close to the dividing line, with no robust way to determine its future evolution considering dynamical relaxation and stellar population ageing. Only by comparing observations with matched simulations can we obtain a good prediction, with the caveat that it is difficult and often ambiguous to connect the kinematics to the merger histories \citep{Naab:2014}.

However, there is also a noticeable velocity gradient in the  north-south direction, that is prolate-like rotation, which exhibits the same sense of motion as the outer gas streams, but with lower amplitude in velocity. The PA of this component is roughly 55\degr\ (counter-clockwise with respect to north, Fig.~\ref{fig:kinematic_maps}) as measured by connecting the highest residual velocity bins. Hence, the stars appear to be a superposition of two different kinematic structures with nearly orthogonal angular momentum vectors. Such a kinematic structure could arise from a polar galaxy merger \citep{Tsatsi:2017} or the spin angular momentum in a radial galaxy merger \citep{Li:2018}.  Hence, a large range of initial merger conditions can lead to prolate-like rotation \citep{Ebrova:2017}, which needs to be explored to infer the particular conditions for NGC~7252 and is beyond of the scope of this paper.

\begin{figure}
   \centering
   \includegraphics[width=\hsize]{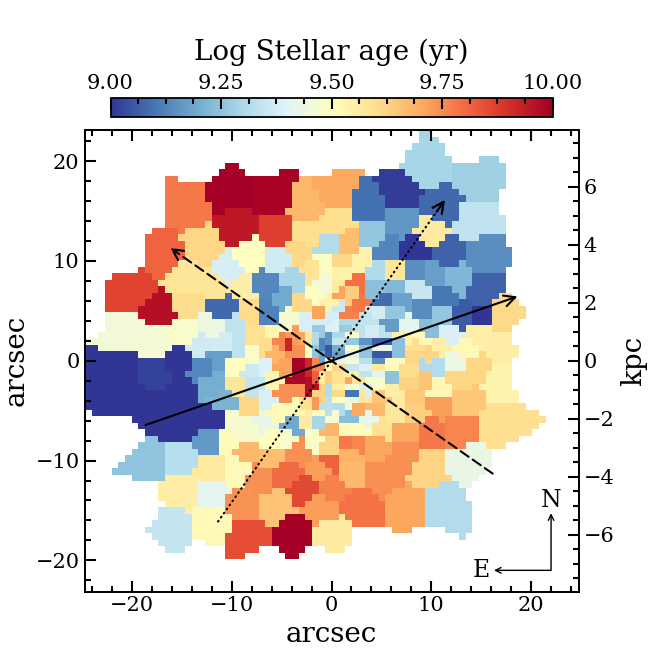}
      \caption{Luminosity-weighted mean stellar ages from \textsc{PyParadise}. Ages are determined from the stellar population templates used to model the stellar continuum from each Voronoi co-added bin. The younger blue population is spatially distinct from the older population in red. Arrows indicating position angles of the primary stellar rotation axis (solid), primary isophotal axis (dashed), and prolate-like rotation (dotted) are borrowed from Fig.~\ref{fig:kinematic_maps}.}
         \label{fig:stellar_ages}
   \end{figure}

This merger configuration is likely reflected in the stellar velocity dispersion map which reveals an elongation spanning some 40\arcsec\ (13 kpc). The primary axis of this high velocity dispersion elongation seems to be orientated nearly along the isophotal semi-major axis within the nearly orthogonal axes of oblate and prolate-like stellar kinematics. Hence, the high velocity dispersion might be primarily caused by the superposition of two different stellar bodies co-existing within intersecting orbital planes. This is qualitatively supported by the structure in the stellar population history as shown later in Fig.~\ref{fig:stellar_ages}.

Another notable feature in the stellar kinematics is an isolated bin just west of the nucleus, which seems to be orbiting counter to the adjacent stars. When compared to photometry, this bin is revealed to be the extensively studied super star cluster W3 \citep[e.g.][]{Maraston:2004, Bastian:2013, Cabrera-Ziri:2016}. The connection between the star formation history of W3 and that of NGC\,7252 will be explored later in section~\ref{SFH}. A similar bin is also found southeast of the nucleus, corresponding to the super star cluster W30.

\subsection{Ionisation conditions} \label{Ionisation_Conditions}
The inner regions of NGC\,7252 are rich in ionised gas, particularly in the central star-forming nuclear star forming disc, the [\ion{O}{iii}] nebulae to the south-west, and a north-easterly gas stream, as shown in Fig.~\ref{fig:kinematic_maps}. The innermost regions show intense H$\beta$ emission, as shown in panel c of Fig.~\ref{fig:kinematic_maps}. Most of the H$\beta$ surface brightness is relatively flat, and concentrated within the star-forming disc. It is also asymmetric, with increased brightness to the south and east. There is slight decrease in the H$\beta$ surface brightness within the central 1\arcsec, which has been previously found in high-resolution narrow-band imaging \citep{Schweizer:2013}. The H$\beta$ distribution falls off quickly towards the edges of the nuclear star forming disc, and remains relatively flat in the annulus immediately outside. H$\beta$ does not, however, feature strongly in the two outer gas streams. 

In contrast, the [\ion{O}{iii}] flux in panel g of Fig.~\ref{fig:kinematic_maps} is centrally concentrated. The distribution falls off much more steeply at all radii, never remaining flat. This may suggest a different ionisation mechanism in the central <$1\arcsec$. [\ion{O}{iii}] does not reach far beyond the nuclear star forming disc, and hence does not feature strongly in the annulus immediately outside. However, the [\ion{O}{iii}] flux is dominant in the SW gas stream.

\begin{figure}
   \centering
   \includegraphics[width=\hsize]{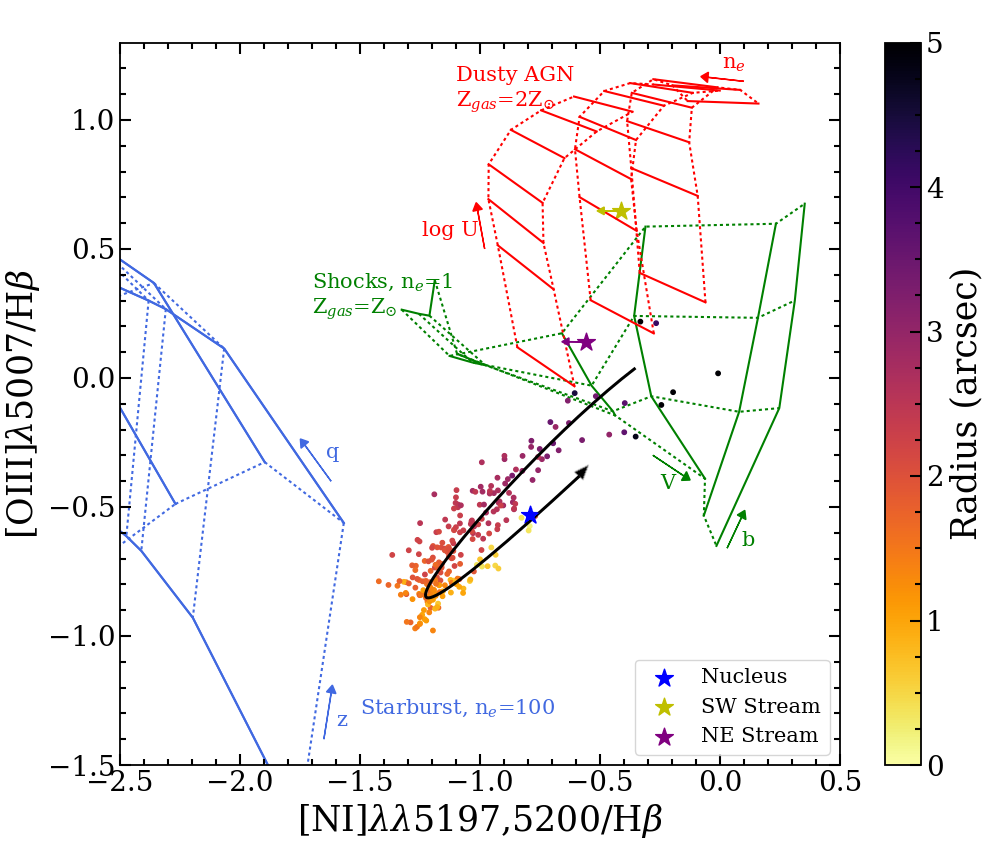}
      \caption{Diagnostic diagram for the inner 5\arcsec\ of NGC 7252. Each point corresponds to a pixel, coloured by radius from the centre of the star forming disc. We highlight the systematic trend of the changing-line ratios with decreasing radius by the curved black arrow. Due to low S/N, we are forced to co-add the outer SW and NE gas streams, and determine an upper limit on \ion{N}{i}. For comparison we highlight model grids based on AGN photoionisation \citep[red,][]{Groves:2004}, shock ionisation \citep[green,][]{Allen:2008}, starburst photoionisation \citep[blue,][]{Levesque:2010} for an instantaneous burst with an age of 1\,Myr.}
         \label{fig:diagnostic}
   \end{figure}

To explore the ionisation conditions of NGC\,7252, we construct a line ratio diagnostic diagram (Fig~\ref{fig:diagnostic}). Since the wavelength coverage of VIMOS prohibits us from constructing the classical [\ion{O}{iii}]/H$\beta$ versus [\ion{N}{ii}]/H$\alpha$ diagram \citep{Baldwin:1981}, we replace [\ion{N}{ii}]/H$\alpha$ with the ([\ion{N}{i}] $\lambda\lambda$5197,5200)/H$\beta$ line ratio. This surrogate line ratio has been exploited successfully by \citet{Sarzi:2010}.

While the S/N for all lines is high enough in individual spaxels within the central 12\arcsec, examining the properties of the extended gas streams requires co-adding spaxels in order to obtain a reasonable S/N. However, even co-added, the spectra lack detection of the weak [\ion{N}{i}] doublet. Hence, we determine 3$\sigma$ upper limits based on the noise of the co-added spectra. The resulting line ratios for the NE and SW gas streams are indicated in Fig~\ref{fig:diagnostic}, respectively.
 
To infer the relevant ionisation mechanisms for each region in the galaxy, we compare the measured line ratios with dusty AGN photoionisation models \citep{Groves:2004}, shock-ionisation models \citep{Allen:2008}, and starburst photoionisation models \citep{Levesque:2010}. The outer regions of the nuclear star forming disc feature shock-driven photoionisation. Moving inwards, we see a smooth transition to towards star formation as the dominant ionisation source. However, the trend reverses again at $\sim$2\,arcsec (0.7\,kpc) at which the gas excitation is increasing again towards shock and AGN ionisation with decreasing distance from the centre. We suspect this is due to an additional ionisation mechanisms at play in the centre of NGC\,7252, possibly due to a low-luminosity AGN, which is discussed in section~\ref{AGN_or_LINER}. At very large radii the co-added emission lines ratios of the NE and SW gas streams both reveal signs of either shocked- or AGN-dominated ionisation, consistent with the bright [\ion{O}{iii}] flux seen in panels g and h of Fig.~\ref{fig:kinematic_maps}. In particular, the high ionization of the SW stream is consistent with the results of the narrow-band imaging of \citet{Schweizer:2013}, who argued that this region may be the light echo of a recent AGN episode.

\subsection{Conditions for ongoing star formation} \label{Conditions_SF}
The nuclear star forming disc is the only site of active star formation in the $50\arcsec \times 50\arcsec$ 
FoV, as evidenced by the aforementioned diagnostic plot shown in Fig.~\ref{fig:diagnostic}. Here, we determine the SFR based on the dust-corrected Balmer line fluxes. While usually H$\alpha$ is employed to count the number of OB stars from their well-known ionising flux, we use the H$\beta$ line converted to the corresponding H$\alpha$ flux. 

\begin{figure}
   \centering
   \includegraphics[width=\hsize]{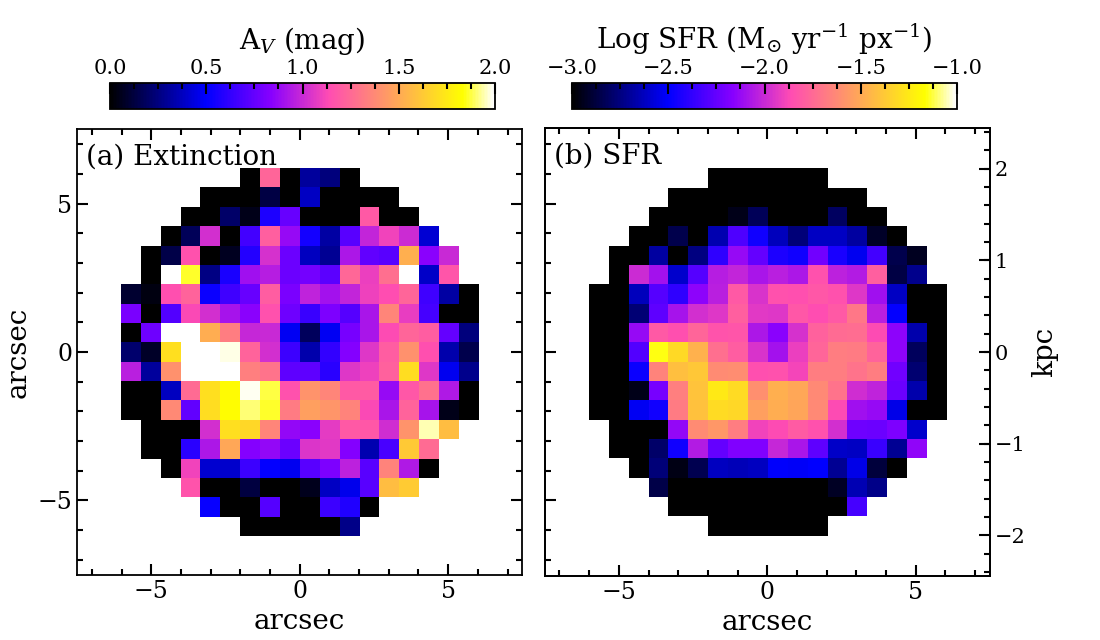}
      \caption{Extinction and SFR maps of the central starburst derived from gas flux measurements with \textsc{PyParadise}. \textit{Panel a} shows the $A_V$ extinction calculated from the observed H$\beta$/H$\gamma$ line ratios based on the Milky Way extinction law \citep{Cardelli:1989}. \textit{Panel b} shows the  estimated map of the SFR per pixel based on the extinction-corrected H$\beta$ luminosity and the prescription of \citet{Calzetti:2013} after conversion to H$\alpha$ luminosity. No signature of ongoing star formation have been identified outside of the central 12\arcsec\ based on emission-line diagnostics.}
         \label{fig:extinction_sfr}
   \end{figure}

We can begin to understand the spatially-resolved dust extinction for NGC\,7252 by examining the H$\beta$/H$\gamma$ line ratio map shown in panel d of Fig.~\ref{fig:kinematic_maps}. Going further, the dust extinction map shown in panel a of Fig.~\ref{fig:extinction_sfr} is derived from the measured H$\gamma$/H$\beta$ line ratio for which we assume an intrinsic line ratio of 0.468 by adopting Case B recombination  with $T_{e}=10^4\,$K and $n_{e}=10^{2} \mathrm{cm}^{-3}$ \citep{Osterbrock:2006}, and the Milky Way extinction law of \citet{Cardelli:1989}. The mean extinction within the central 12\arcsec\ region is found to be 0.65\,mag.

In examining both the Balmer ratio and the dust extinction map, one may note the apparent ring structure first seen in studies of the ionised gas by \citet{Schweizer:1982} and then of the molecular gas by \citet{Wang:1992}. We also note a relatively high extinction region in the western portion of the ring, which corresponds to dust features seen in optical imaging, as well as the asymmetric H$\beta$ distribution mentioned in section~\ref{Modelling}. 

To use H$\alpha$ as a SFR calibration, we first compute the dust-corrected H$\alpha$ flux from H$\beta$ adopting an intrinsic H$\alpha$/H$\beta$ line ratio of 2.863 which is based consistently on Case B conditions with T$_{e}=10^4$\,K and $n_{e}=10^{2}\mathrm{cm}^{-3}$ \citep{Osterbrock:2006}. This leads to an integrated dust-corrected H$\alpha$ luminosity of  $L_{\mathrm{H}\alpha} = (3.9\pm0.4)\times10^{41}\mathrm{erg~s}^{-1}$ from which we then calculate the SFR using the prescription of \citet{Calzetti:2013} given by 
\begin{equation} \label{equ:CAL_SFR}
\mathrm{SFR}(\mathrm{H}\alpha)/[M_{\sun}\mathrm{yr}^{-1}] = 5.5\times 10^{-42} L_{\mathrm{H}\alpha}(\mathrm{erg~s}^{-1}),
\end{equation}
which assumes a Kroupa initial mass function \citep{Kroupa:2001} and Case B conditions consistent with our dust correction procedure. The resulting star formation rate map is shown in panel b of Fig.~\ref{fig:extinction_sfr} which is clipped to show only pixels predominately ionised by star formation, as indicated by the diagnostic plot in Fig.~\ref{fig:diagnostic}. The corresponding SFR within the VIMOS FoV is computed to be $\mathrm{SFR}(\mathrm{H}\alpha)=2.2\pm0.6\,M_{\odot}\,\mathrm{yr}^{-1}$. Since star formation is also known to happen far away from the galaxy centre in the tidal tails of NGC~7252 \citep[e.g.][]{Lelli:2015},  we also estimated the total SFR. FIR fluxes were measured by \textit{AKARI} \citep{Kawada:2007} which are $f_{90\mu\mathrm{m}}=5.1$\,Jy and $f_{140\mu\mathrm{m}}=6.4$\,Jy as listed in the point source catalogue \citep{Yamamura:2010}. Following the prescription of \citet{Takeuchi:2010} we compute an FIR-based SFR of $\mathrm{SFR(FIR)}=6.7\,M_{\odot}\,\mathrm{yr}^{-1}$ also for a Kroupa IMF. Hence,  a substantial fraction of the total star formation is confined to the central few kpc.

CO(1-0) measurements of NGC 7252 obtained by ALMA have provided a molecular gas mass of $4.3 \times 10^{9}$ M$_{\odot}$ for the inner gas disc \citep{Ueda:2014}. Hence, we calculate a molecular depletion time of $t_\mathrm{dep}=1.9\pm0.6$\,Gyr. This is surprisingly close to the average molecular depletion time of 2.35\,Gyr for normal star forming disc galaxies \citep{Bigiel:2011}, but shorter than expected from quiescent elliptical galaxies \citep[e.g.][]{Saintonge:2012}. While gas-rich early-type galaxies \citep{Davis:2014} exhibit on average a factor of $\sim$$2.5$ lower depletion times, the depletion time can be as short as for normal galaxies if the molecular gas is located more the flat part of galaxies rotation curve.

\begin{figure}
   \centering
   \includegraphics[width=\hsize]{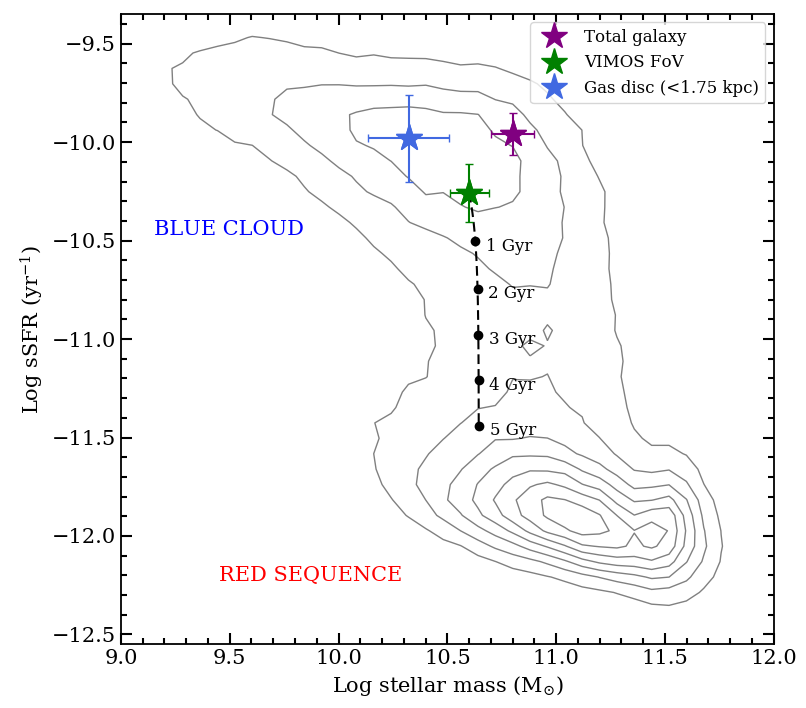}
      \caption{Specific SFR (sSFR) against stellar mass for NGC~7252 corresponding to the total galaxy (purple), VIMOS field-of-view (green), and nuclear star forming disc (blue). The bimodal galaxy distribution from SDSS MPA-JHU catalog \citep[][]{Kauffmann:2003b,Brinchmann:2004} is shown as contours for comparison. Assuming a constant depletion timescale of $t_\mathrm{dep}=1.9\pm0.6$\,Gyr for NGC\,7252 based on the molecular gas mass derived from CO(1-0) \citep{Ueda:2014} we predict the position of NGC\,7252 for the next 5\,Gyr in case of simple gas consumption.  
              }
         \label{fig:evolution}
   \end{figure}

Based on the results above we explore the location of NGC\,7252 in the sSFR--$M_{*}$ plane (Fig.~\ref{fig:evolution}). We use the archival \textit{Hubble} WFC3 images of NGC\,7252 in the F475W (Sloan $g$) and F775W (Sloan $i$) filters to infer a brightness of $m_g=12.56$\,mag (AB) and $m_i=11.80$\,mag (AB), respectively, within an aperture of $30\arcsec$ radius. Following the empirical stellar mass calibration of \citet{Taylor:2011}, 
\begin{equation}
 \log(M_*/[M_\sun]) = 1.15 +0.7(g-i) - 0.4 M_i
\end{equation}
we compute a stellar mass for NGC\,7252 of $\log(M_*/[M_\sun]) = 10.6\pm0.1$. Compared to the total stellar mass through a $80\arcsec$ radius aperture, the VIMOS FoV actually covers 70\% of the stellar mass. Together with the SFR computed for the VIMOS FoV and total galaxy as described above, we compute  specific star formation rates (sSFRs) of $(5.4\pm1.9)\times$10$^{-11}$ yr$^{-1}$ and $(11\pm3)\times$10$^{-11}$ yr$^{-1}$ for the VIMOS FoV and the total galaxy, respectively. Comparing this to the distribution of local galaxies from the SDSS MPA/JHU galaxy catalogs \citep[e.g.][]{Kauffmann:2003b,Brinchmann:2004} we find that, NGC\,7252 is still located in the blue cloud of star forming galaxies. We note that the region of ongoing star formation is much smaller than the host galaxy. Hence, we calculate the sSFR using the dynamical mass of the nuclear star forming disc as described in section~\ref{gas_kinematics}, which is marked in Fig.~\ref{fig:evolution}. Since the mass of the star forming disc is smaller than the total mass by a factor of three within the VIMOS FoV, the sSFR of this region is right on the main sequence of star formation. This suggests that the galaxy still contains a settled disc component, which has strong implications for the future evolution of NGC\,7252 as discussed later in section~\ref{Destiny}.

\subsection{Star formation history} \label{SFH}

   \begin{figure}
   \centering
   \includegraphics[width=\hsize]{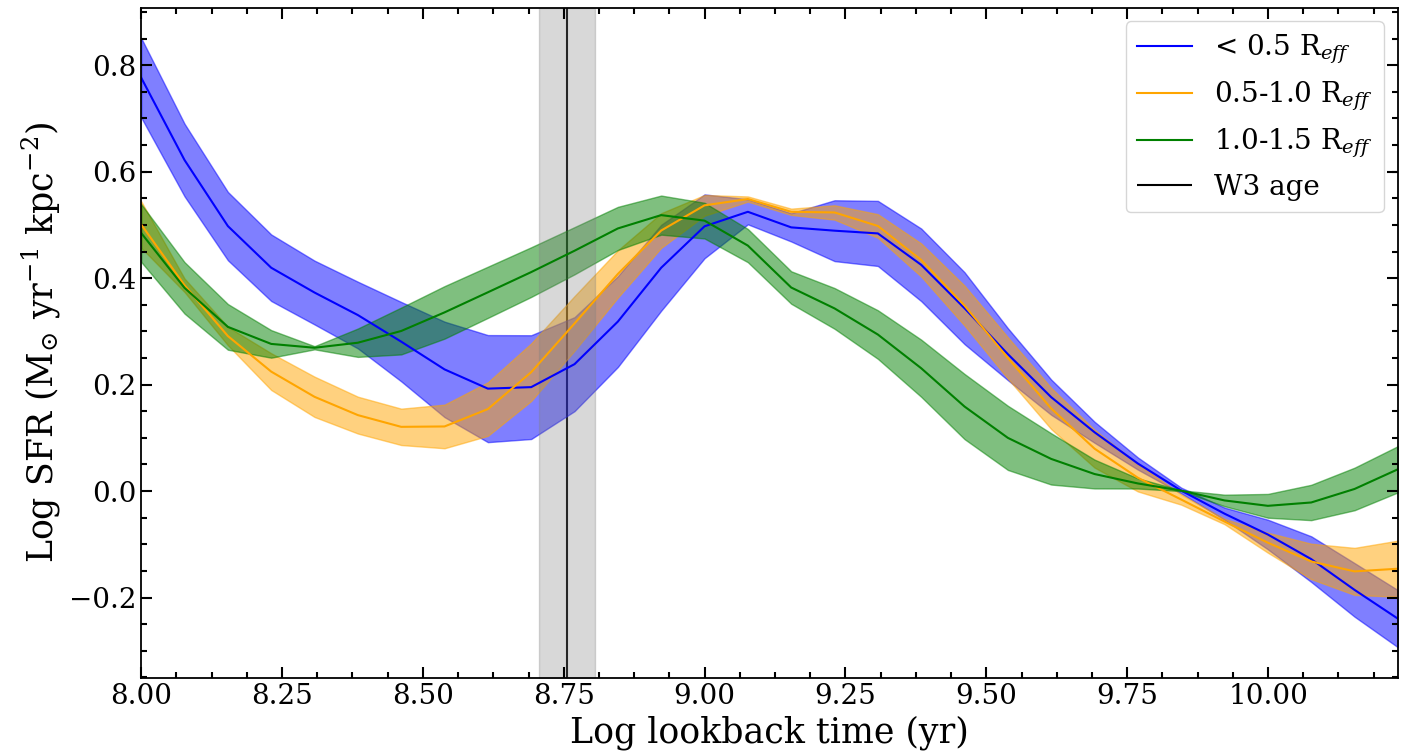}
      \caption{Star formation histories binned by effective radius. Each curve is binned by 0.5 effective radius, with errors determined from the standard deviation within each spatial bin at a given age. The black vertical line indicates the age of the associated super star cluster W3 as reported by \citet{Cabrera-Ziri:2016}, with the uncertainty shown in grey.}
         \label{fig:sfr_curves}
   \end{figure}  

   \begin{figure*}
      \resizebox{\hsize}{!}{\includegraphics{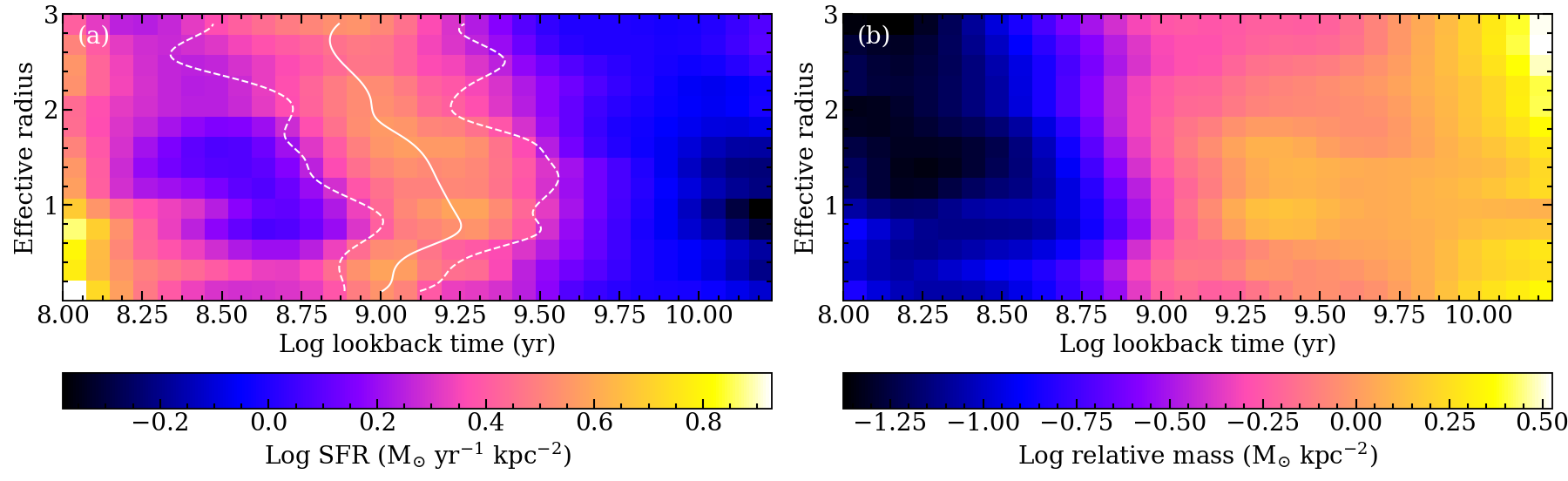}}
      \caption{Summary of the star formation history of NGC\,7252 obtained with STECKMAP, normalised over each annulus. \textit{Panel a} shows the SFR as a function of effective radius, measured across cosmic time. The white solid and dotted lines outline the mean and standard deviation of the initial star formation episode. The current star formation episode is clearly visible in the lower left-hand corner. Similarly, \textit{panel b} shows the stellar mass gained at a given time. As expected, most of the mass was built up slowly, beginning at early times with a second peak around 1\,Gyr.} 
      \label{fig:sfh_colormap}
   \end{figure*}
     
We use the STEllar Content and Kinematics from high-resolution galactic spectra via Maximum A Posteriori code \citep[STECKMAP,][]{Ocvirk:2006a,Ocvirk:2006b} to measure star formation histories (SFHs) across the VIMOS FoV. STECKMAP recovers age and metallicity distributions in a non-parametric way, by fitting the full observed spectrum. In practice, STECKMAP minimizes the objective function $Q_\mu$ defined as
\begin{equation}
 Q_\mu = \chi^2(\mathbf{x},\mathbf{Z},\mathbf{g}) + P_\mu(\mathbf{x},\mathbf{Z},\mathbf{g})
\end{equation}

\noindent where $\mathbf{x}$ and $\mathbf{Z}$ are the age and metallicity distributions, and $\mathbf{g}$ is the discretised line of sight velocity distribution (LOSVD). The term $P_\mu$ is the penalisation function
\begin{equation}
 P_\mu = \mu_\mathbf{x} P(\mathbf{x}) + \mu_\mathbf{Z} P(\mathbf{Z}) + \mu_\mathbf{v} P(\mathbf{g})
\end{equation}
 
\noindent This penalisation favours smooth solutions and, effectively, it acts as a Gaussian prior $f_\mathrm{prior} = e^{-\mu P}$ \citep{Ocvirk:2006a}. Notice that the penalisation coefficients $\mathbf{\mu}$ are set separately for the age, metallicity, and LOSVD distributions. 

Although STECKMAP allows for a simultaneous recovery of both stellar populations and kinematical properties, it has been shown that metallicity and stellar velocity dispersion are degenerate parameters \citep{Koleva:2008}. However, this degeneracy can be safely broken by first measuring the discretised LOSVD, and then fixing it while measuring the stellar population properties \citep{Sanchez-Blazquez:2011}. The LOSVD measurements are already performed with \textsc{PyParadise} as described in section~\ref{stellar_kin} and used as inputs for STECKMAP.

We feed STECKMAP with the MILES stellar populations synthesis models \citep{Vazdekis:2010}. This set of models ranges from $-1.3$ to $-0.22$ dex in metallicity, and from 67 Myr to 17 Gyr in age. We note that the MILES base models follow the solar neighbourhood abundance pattern, meaning that they are [Mg/Fe]-enhanced at low metallicities, but [Mg/Fe] $\sim0$ for [Fe/H] $\gtrsim 0$. We keep the stellar initial mass function fixed to that measured in the Milky Way \citep{Kroupa:2001}. Since the S/N drops significantly on the blue end of the spectra, we limit our analysis to wavelengths between $\lambda = 4500$\AA\ and  $\lambda = 5500$\AA. For this work, we assume $\mu_\mathbf{x} = \mu_\mathbf{Z} = 20$. This choice is motivated by a joint comparison between the minimum $\chi^2$ value and the structure of the SFHs. Large $\mu$ values lead to smoother solutions, therefore losing time-resolution, but assuming low $\mu$ values would produce artefacts on the recovered age and metallicity distributions. The choice of a reliable $\mu$ vector is not trivial, but does not affect the overall shape of the SFHs. We have assessed this point by varying the adopted $\mu$ values, finding no significant differences in our results.

Finally, the age distributions recovered using STECKMAP are translated to mass fractions and SFRs as a function of look-back time, as demonstrated in Fig.~\ref{fig:sfr_curves} and Fig.~\ref{fig:sfh_colormap}. The recovered SFHs strengthen the story of NGC\,7252 as a merger remnant. Fig.~\ref{fig:sfr_curves} demonstrates that the SFR peaked roughly 1 Gyr ago whereby the burst occurred slightly earlier in the innermost region (blue) than in the outer regions (orange, then green) at a similar SFR. The current episode of rapid star formation is significantly stronger in the inner region. 

Not only does the merger age estimate of 0.6 Gyr as suggested by \citet{Schweizer:1982} fit well into this picture, but also the estimated age of 570\,Myr for star cluster W3 as inferred by \citet{Cabrera-Ziri:2016}. In contrast, the star formation history of the entire galaxy appears significantly extended over several 100\,Myr during the merger period, more than just a single burst. We will discuss the relation of this extended star formation with its spatial distribution as a result of the merger in the following section.
   
\section{Discussion} \label{Discussion}
\subsection{Prolate-like rotation from a gas-rich major merger}
Early-type galaxies are usually rotation around their minor axis (oblate rotation), if they regularly rotate at all. Evidence for systems with rotation around the  major axis (prolate-like rotation) instead is currently scarce. After many unsuccessful searches \citep[e.g.][]{Bertola:1988}, there are currently about a dozen early-type galaxies known to have clear signatures of prolate-like rotation \citep[e.g.][]{Schechter:1979,Wagner:1988, Krajnovic:2011, Emsellem:2014}. IFU surveys like CALIFA \citep{Sanchez:2012a}, MaNGA \citep{Bundy:2015} or the Most Massive MUSE Galaxies (M3G) survey (PIs: Emsellem \& Krajnovi\'c) are working to increase the number of galaxies with signatures of prolate-like rotation.

The nature and origin of the prolate-like rotation in early-type galaxies is still under debate. Satellite accretion, as well as major mergers of nearly equal mass have become the two leading formation scenarios \citep{Ebrova:2017, Li:2018}. \citet{Tsatsi:2017} has discovered ten early-type galaxies in CALIFA with signatures of prolate-like rotation, which corresponds to 9\% of all early-type galaxies in the same volume. By comparing the observations with N-body simulations they showed that the prolate-like rotation can be produced by a major polar merger.  As NGC\,7252 has formed from the merger of two disc galaxies, it is very likely that its formation mechanism and observed prolate-like rotation are indeed related. We therefore expect that some part of the angular momentum vectors, composed of the orbital and rotational vectors, should be perpendicular to each other. Whether the prolate-like rotation is predominately produced by one of those angular momentum components is unclear at this point. However, the vast majority of presently known galaxies with prolate-like rotation are found in dense environments and have little or no gas. NGC~7252 is in both ways different from those, suggesting its formation mechanism is different than other known cases. Alternatively, with NGC~7252 we are witnessing the early, still gas rich, stage of the formation of a future classical elliptical with complex kinematics.

Binary mergers are known to produce galaxies with prolate-like rotation \citep[e.g.][]{Naab:2003}, where the actual kinematics, such as the existence of a kinematically decoupled core (KDC), or the change from an oblate-like (centre) to a prolate-like rotation (outskirts) is dependent on the mergers orbit, spins of the progenitors and the gas content \citep{Hoffman:2010, Bois:2011}. NGC\,7252 still has a considerable amount of gas, $4.5\times10^9M_\sun$ of \ion{H}{i} \citep{Hibbard:1994} and $3.5\times10^9M_\sun$ of $H_2$ \citep{Wang:1992}, and its kinematic structure could be viewed as an inner KDC exhibiting an oblate-like (regular) rotation, which changes beyond 10\arcsec\ (3.3 kpc) into a prolate-like rotation. The velocity map of NGC\,7252 resembles those of binary merger simulations of \citet{Hoffman:2010} with gas fractions of 20\% or more. In these simulations, the gas is responsible for building the short-axis tubes and the regular rotation in the centre, while the outer prolate-like rotation (built of long-axis tubes) can appear in various combinations of the orbital parameters and progenitor spins, and does not necessarily require a polar orbit. Similar conclusions are also found in \citet{Bois:2011}.

The first numerical simulations of NGC\,7252 by \citet{Borne:1991} and \citet{Mihos:1993} employed a retrograde merger of disc galaxies to account for the presumably counter-rotating gas disc reported by \citet{Schweizer:1982}. Later, \citet{Hibbard:1995} presented a new self-consistent $N$-body simulation based solely on the large-scale H\,I kinematics obtained with the Very Large Array. In this model, the inclination of two progenitor galaxies are $i_1=-40$\degr\ and $i_2=70$\degr\ with respect to the merger plane, which would imply perpendicular orbital angular momentum vectors, consistent with a polar merger geometry. However, various other configurations may also be able to produce the observed stellar kinematics that need to be systematically tested by varying the initial conditions of this merger in terms of mass ratio, impact parameter, orbital and spin angular momentum. As stars are collisionless particles, in contrast to the gas, they can probe the kinematics down to the centre of the remnant galaxy, void of \ion{H}{i} gas. While performing matched $N$-body simulations is beyond the scope of this paper, our stellar kinematic field should provide valuable constraints on merger parameters for future hydrodynamical simulations.

\subsection{The star formation history during the merger event} \label{Evolution}
NGC\,7252 provides a snapshot in the evolution of two spiral galaxies into an elliptical. Dedicated numerical simulations of this particular merger remnant have constrained the time since the pericentre of the two discs to be about 770\,Myr in the simulations by \citet{Borne:1991} and 620\,Myr in the simulations by \citet{Chien:2010}. This timescale is supported by the ages of associated massive star clusters lying in the range of 400-600\,Myr \citep{Schweizer:1998}. The age of the well-studied cluster W3, for example, has been found at 570\,Myr \citep{Cabrera-Ziri:2016}. The centre of NGC\,7252 exhibits a post-starburst spectrum with strong Balmer lines \citep{FritzeAlvensleben:1994}, implying a strong starburst event less than 1\,Gyr ago and an extended period of star formation starting even before pericentre passage. With our IFU data we are able to explore the star formation history of NGC\,7252 for the first time, elucidating the nature of the intense star formation episode coincident with the merger.

One main result of our IFU observation is that the starburst episode is significantly extended over several hundred Myr and suggests star formation activity before, during, and after the time of first pericentre passage. In particular, we show that the peak in star formation activity first occurred near the galaxy centre ($<1R_\mathrm{eff}$), moving radially outwards at later times. The extended star formation activity prior to the anticipated time of pericentre passage could be explained by a starburst which had been already triggered in the galaxy core, as the galaxies were approaching the closest point of their first pass. Such enhanced star formation activity at the centre of galaxies has already been measured over large samples of galaxies at separations of $\lesssim$20\,kpc \citep{Li:2008b,Ellison:2008,Wong:2011,Scudder:2012,Patton:2013, Davies:2015}. Hence, it is observationally expected that star formation is already enhanced before the final pericentre passage. This is consistent with numerical simulations of nearly equal mass mergers of disc galaxies, exhibiting the highest SFR excess a few hundred Myr before the final pericentre passage \citep[e.g.][]{Moreno:2015} and a suppression after final coalescence.

Our spatially-resolved SFH also recovers the period of low star formation activity after pericentre, which reaches a minimum around $\sim$300\,Myr ago. The decline in SFR is less effective for the regions which end up at larger radii possibly due to  a higher angular momentum than the remnant centre. Those regions may therefore originate primarily from the outskirts of the progenitors rather than the cores and may subject to continuous or enhanced star formation at later time during the merger. The relatively young ages and galacto-centric radius of the star clusters such as W3 matches well with the extended episode of star formation in the outskirts of the merging system. Such a shift in SFR peak from the core to the large scale regions is also reproduced in the numerical simulations by \citet{Moreno:2015}.

Although the nucleus of NGC\,7252 exhibits post-starburst features, ongoing star formation in the central 2\,kpc is clearly present. This is readily seen in the current epoch by an increased SFR at small radii, with a similar SFR level appearing to have existed previously $\sim$\,1 Gyr ago. This on-going star formation activity suggest that the final coalescence of the major merger occurred relatively recently, certainly within the last 200 Myr: numerical simulations of major mergers with masses and dynamical times show that merger-induced star formation stops within 100--200 Myr after the coalescence, even without AGN quenching \citep{Springel:2005, Bournaud:2011, Torrey:2012, Powell:2013, Moreno:2015}. High-resolution hydrodynamic simulations resolving dense gas clouds within merging galaxies have shown that triggered star formation can last longer than previously thought, but only before the coalescence, and rapidly decreases in 100--200 Myr afterwards \citep[e.g.][]{Teyssier:2010, Renaud:2014}. Observations also suggest that merger-induced star formation events are short-lived \citep{Ellison:2013}.

This also holds for systems at high redshift and/or with high gas fractions \citep{Fensch:2017}. Dedicated simulations fitting NGC\,7252 by \citet{Chien:2010} also predict rapidly falling SFR after the starburst phases. In detail these simulations do not predict any recent star formation activity, but their resolution limited to 170\,pc for the gravitational softening with a potentially larger SPH smoothing length may miss local events in the nucleus itself – alternatively the true orbit, structure, and kinematics of the progenitor models which merged to form NGC\,7252 may somewhat differ from those used in this specific simulation.

Another line of evidence for recent coalescence comes from the bright tidal features visible around NGC7252: such features rarely last more than $\sim$\,500 Myr \citep[e.g.][]{Duc:2013}. The study of tidal debris in post-merger systems by \citet{Ji:2014} indicates that such debris are likely visible at magnitudes brighter than 25\,$\mathrm{mag}\,\mathrm{arcsec}^{-2}$ during 600 Myr or less for systems with the mass of NGC\,7252: debris in NGC\,7252 is both much brighter than this limit and fairly continuous over tens of kpc, suggesting that the previous value is a strong upper limit to the age of this post-merger system.

These lines of evidence show that the presumably re-formed disc at the centre of NGC\,7252 is young, most likely 100-200 Myr-old. Such rapid re-formation of a relaxed and symmetric cold gas disc is consistent with high-resolution simulations of major mergers \citep[e.g.][]{Bournaud:2011, Renaud:2014}. The settled gas must have been located in-situ in or near the galaxy nuclei or at least in the inner progenitor galaxy discs, as accretion from the outskirts would require much longer timescales.

Although the SFH of NGC\,7252 agree well with generic predictions of merger simulations, some aspects remain puzzling. We anticipate that the spatially-resolved SFH of NGC\,7252 will provide useful constraints on the initial conditions of gas content and distribution for tailored simulations of this particular major merger system.

\subsection{Low-luminosity AGN or LINER?} \label{AGN_or_LINER}
One of the most recent discoveries from the study of NGC\,7252 is the detection of a bright [\ion{O}{iii}] nebulosity $\sim$5\,kpc south-west of the galaxy centre by \citet{Schweizer:2013}. No indication of ongoing black hole (BH) activity at the nucleus has been yet confirmed in the radio and X-rays, and so \citet{Schweizer:2013} concluded that NGC\,7252 may have hosted an AGN at its centre that has drastically decreased in luminosity in the recent past. NGC\,7252 would therefore fall in the same category as Hanny's Voorwerp \citep{Lintott:2009} and other galaxies that display potential AGN light echos \citep{Keel:2015}. Our IFU observations also cover this [\ion{O}{iii}] nebulosity and we confirm the high excitation of the gas. Based on our emission-line diagnostic diagrams as discussed in section~\ref{Ionisation_Conditions}, we also find evidence of an AGN as the dominant ionisation mechanism in this particular region.

In addition, our VIMOS IFU data provides additional constraints on the gas excitation around the nucleus. As shown in Fig.~\ref{fig:extinction_sfr}, the dust attenuation map and SFR distribution reveal a drop in dust content and H$\alpha$ luminosity at the galaxy centre. This agrees with a previously reported deficit within the central 1-2\arcsec\ of H$\alpha$ from narrow-band images \citep{Schweizer:1982} and molecular gas \citep{Wang:1992}. However, the centrally concentrated [\ion{O}{iii}] emission leads to an enhanced [\ion{O}{iii}]/H$\beta$ line ratio. The emission-line diagnostic plot (Fig.~\ref{fig:diagnostic}) confirms that the line ratios indeed increase towards higher excitation at the galaxy centre. \citet{Schweizer:2013} also compared the line ratios from narrow-slit spectra of the galaxy centre, but only considered a nuclear and a star forming disc size aperture, missing the gradient apparent in the emission profiles. Given the low spatial resolution of our data we cannot avoid a strong contamination of the nuclear emission lines by the bright star-forming disc, in particular for H$\beta$. The intrinsic line ratios of the nucleus are then likely located in either the shock or the low-luminosity AGN region of the diagram. Only IFU observations at higher spatial resolution may be able to spatially separate the line ratios from the nucleus and the star forming disc.

If we nevertheless interpret the extinction-corrected ($A_V \sim 1.35$\,mag) [\ion{O}{iii}] line flux of $(3.17\pm0.3)\times10^{-15}\,\mathrm{erg/s/cm}^2$ within the central $4\arcsec\times4\arcsec$ to be excited by a low-luminosity AGN, we would estimate an AGN bolometric luminosity of $L_\mathrm{bol}=(2.4\pm0.3)\times10^{41}\,\mathrm{erg~s}^{-1}$ following the empirical correlation between [\ion{O}{iii}] and bolometric luminosity from \citet{Stern:2012b}. The corresponding 2-10\,keV luminosity would be $5.9\times10^{40}\,\mathrm{erg~s}^{-1}$, adopting a bolometric correction of $L_\mathrm{[\ion{O}{iii}]}/L_\mathrm{2-10keV}\sim0.03$ from \citet{Heckman:2005}. This is perfectly consistent with the X-ray luminosity of the nucleus $\log(L_X/\mathrm{[erg~s^{-1}]})=40.75^{+0.03}_{-0.04}$ as reported by \citet{Nolan:2004} from \textit{Chandra} observations. Hence, the nuclear [\ion{O}{iii}] and X-ray emission indeed follows the relation implied for BH activity. \citet{Schweizer:2013} concluded that most of the X-ray luminosity may originate from star formation from the circumnuclear disc. However, they assume a SFR of $6$\,$M_\sun\mathrm{yr}^{-1}$ for the star-forming disc, which is a factor of three higher than we infer from the IFU data as discussed in section~\ref{Conditions_SF}. 

In any case, the putative AGN luminosity is still much lower than the luminosity required to excite the [\ion{O}{iii}] nebulosity as discussed by \citet{Schweizer:2013} and their scenario of a recently faded AGN remains valid. These kind of changing-look AGN are observed with increasing numbers \citep[e.g.][]{Denney:2014,LaMassa:2015,Merloni:2015,McElroy:2016}. In particular, Mrk\,1018 is not only a galaxy in an advanced major merger stage, but also shows AGN flickering on a timescale of a few years. This flickering may possibly be related to the interaction of a binary BH system, accretion disc wind regulation, or other currently unknown scenarios \citep{Husemann:2016b}. Whether the variability in BH accretion on $10^3-10^5$yrs timescales is particularly enhanced during particular phases of a major merger or simply reflects the usual variability of AGN \citep[e.g.][]{Schawinski:2015} remains to be tested.

Alternatively, the emission-line ratios may also point to a Low-Ionisation Nuclear Emission Region (LINER) for which AGN are not necessarily the primary ionisation mechanism. Also, shocks from supernova or the gravitational infall of gas would lead to line ratios consistent with slow shocks. Furthermore, post-AGB stars have become a more favourable interpretation of LINERs in early-type galaxies \citep[e.g.][]{Singh:2013}. The high central concentration of [\ion{O}{iii}] emission also agrees with such a scenario given the steep surface brightness profile of the stellar continuum corresponding to a de~Vaucouleurs law \citep[e.g.][]{Schweizer:1982}. Again, given our poor spatial resolution, we cannot readily distinguish between a pure point-like or slightly extended [\ion{O}{iii}] flux distribution. Also, the contamination of H$\beta$ due to the star forming ring is too high to test whether the expected amount of ionising photons from post-AGB stars is sufficient to explain the excitation, as we are using the equivalent width of H$\beta$ \citep[as a surrogate for H$\alpha$;][]{CidFernandes:2010}. In this case, the estimated AGN bolometric luminosity from the [\ion{O}{iii}] line would be a very strong upper limit, strengthening the fading AGN scenario.

\subsection{When will NGC~7252 reach the red sequence?} \label{Destiny}
While NGC\,7252 is still located within the blue cloud of star-forming galaxies in colour-magnitude and sSFR-$M_*$ plane, the high S\'ersic index of the merger remnant is consistent with the standard evolutionary picture in which the major merger of two star-forming disc galaxies is responsible for the morphological transformation to an elliptical galaxy. Also, a high SFR is predicted to occur at several times during the merger, as discussed above. However, the process of transitioning from a blue cloud to a red sequence galaxy is less clear. On which timescales the stellar populations evolve is a key question, being either a fast or a slow process. The relatively low number density of green valley galaxies implies that the transition phase should be short, lasting much less than 1\,Gyr \citep[e.g.][]{Schawinski:2007, Schawinski:2014}. On the contrary, \citet{Martin:2007} found a much longer timescale of 1-6\,Gyr for the transition from the blue cloud to the red sequence. Only the latter scenario appears consistent with the gas consumption timescale, which may be the main regulator for the future evolution of NGC\,7252. This implies that the transition phase for NGC\,7252 is either significantly prolonged and unexpected for gas-rich major mergers, or that other processes such as AGN feedback have yet to kick-in. The later must occur in order to support a rapid quenching of star formation \citep[e.g.][]{Springel:2005,Matteo:2005}.

The long transformation time we infer is based on the assumption that the $t_\mathrm{dep}$ remains unchanged until all the molecular gas is consumed. This timescale may be significantly shortened in the event of another burst in star formation, an increasing fraction of the molecular gas being locked up in the diffuse atomic gas phase and not able to form stars, dynamical suppression of star formation or an expulsion of gas by an AGN. The first option is less likely given that the merger has already passed the coalescence phase and no additional triggering of an enhanced star formation phase is expected in numerical simulations \citep[e.g.][]{Chien:2010}. In line with the second option, \ion{H}{i}-rich elliptical galaxies have been detected and represent roughly 40\% of the local ellipticals \citep{Serra:2012} outside of clusters,  so it is possible that the final remnant of NGC\,7252 will be a \ion{H}{i}-rich elliptical galaxy. The dynamical suppression of star formation in early-type galaxies was put forward by \citet{Davis:2014}, who found that the depletion time was significantly longer for molecular gas located in the rising part of the rotation curve. This mechanisms seem not to apply in the case of NGC\,7252 because most of the molecular gas in the star-forming disc is located flat part of the rotation curve as traced by H$\alpha$ emission (see Fig.~\ref{fig:velocity_curve}). Hence, additional tidal forces would be required to bring the gas closer in towards the nucleus on short timescale, which is unlikely to be happen given the advanced evolutionary stage of the merger.

AGN triggering during a major merger is often invoked to rapidly quench star formation by expelling much of the gas with powerful outflows \citep[e.g.][]{Matteo:2005}. The strong [\ion{O}{iii}] nebulosity of NGC\,7252 already shows that an AGN was likely present in the recent past, as convincingly argued in \citet{Schweizer:2013}. This is also consistent with the presumed time delay of a few 100\,Myr between the peak of star formation and AGN accretion \citep[e.g.][]{Wild:2010,Schawinski:2010b}. However, no kinematic signature of an AGN-driven outflow on galaxy-scales has yet been detected for NGC~7252. Apparently, the previous AGN phase was either too brief or too weak to develop a galactic outflow. It is unknown when the next AGN phase will be triggered and if it will be powerful enough to develop a sufficiently strong outflow to rapidly quench star formation. The IFU data we present is consistent with the presence of a low-luminosity AGN and an increasing number of AGN are known to change their luminosity on rather short timescales \citep[e.g.][]{Lintott:2009,Denney:2014,LaMassa:2015,McElroy:2016}. Even if an AGN will turn on again, it may not be able to affect the settled gas disc, as some simulations have shown that the AGN energy may be able to escape without removing much gas in this configuration \citep[e.g.][]{Gabor:2014}. Hence, NGC~7252 already passed the presumed AGN blow-out phase for a purely merger-driven AGN quenching scenario. If the modest gas reservoir currently available to form stars is not first removed by an AGN phase or rapidly depleted by a burst of star formation triggered by dynamical effects (e.g. minor mergers), we conclude that NGC\,7252 will eventually die out and become a red elliptical galaxy in several Gyr as predicted for slow transformation scenario \citep[e.g.][]{Martin:2007}.

\section{Summary and conclusions} \label{Conclusions}
Using VLT-VIMOS IFU observations covering the central $50\arcsec\times50\arcsec$ of the recent merger remnant NGC\,7252, we obtain maps detailing the stellar and gaseous properties of the galaxy, including kinematics and line strengths. Our major results based on a detailed analysis of these spatially-resolved properties can be summarised by the following:
\begin{itemize}
 \item We confirm the presence of a central rotating gas disc with a kinematic inclination and maximum velocity in agreement with \citet{Ueda:2014} and no signature of counter-rotation as initially reported. 
 \item We find clear signatures of both oblate- and prolate-like rotation which may be caused by a polar merger configuration or a radial orbit with different spin angular momentum vectors.
 \item A study of the SFH of NGC\,7252 indicates an extended period of elevated SFR extending until 1\,Gyr ago, reaching beyond the anticipated time of first passage, which highlights a complex radially-dependent SFH produced throughout the galaxy merger.
 \item We compute a SFR of $2.2\pm0.6M_\sun\mathrm{yr}^{-1}$ as measured by the dust-corrected H$\alpha$ luminosity within the central 7\,kpc, corresponding to a molecular gas depletion time of $\sim$2\,Gyr given the measured molecular gas mass in this region. 
 \item Higher ionisation conditions of the gas are found at the very centre of NGC\,7252 from which we draw an upper limit on $L_\mathrm{bol}<2.4\times10^{41}\,\mathrm{erg~s^{-1}}$ based on the [\ion{O}{iii}] luminosity, consistent with the X-ray luminosity of the nucleus.
 \item NGC~7252 is still within the blue cloud of galaxies as seen in the sSFR--$M_*$ plane and assuming a constant molecular gas depletion time we infer a transitional time of about 5\,Gyr until it reaches the red sequence through gas consumption.
 
\end{itemize}

Given that NGC\,7252 is one of the nearest major merger remnants of two disc galaxies, it has been employed as a Rosetta stone in understanding the transformational process from star-forming disc galaxies to quiescent elliptical galaxies. Our IFU observations provide a suite of new diagnostics such as prolate-like rotation in the stellar components and a complex spatially-resolved star formation history. All of this information needs to be interpreted alongside dedicated numerical simulations of major mergers, which can now be better tuned to NGC\,7252, beyond the scope of this work. 

The past history of NGC\,7252 seems more and more settled as simulations and observations are converging to a common picture. The final destiny however, still remains open. While NGC\,7252 has already transformed into an early-type galaxy based on the light profile, it remains as a blue cloud galaxy based on colours and SFR. A detailed analysis of the star formation history and tidal debris suggests that the central gas disc has presumably re-formed quickly in within 100 Myr since central coalescence of the major merger, most likely from in-situ material rather than infall from the galaxy outskirts. Hence, the transformation from two blue disc galaxies to a quiescent elliptical is far from complete as a significant portion of the transformation has not yet begun. It seems the destiny of NGC\,7252 cannot be constrained nor predicted at this point. The most plausible scenario is that the galaxy will likely evolve on the timescale of several Gyr as it slowly consumes the remaining gas. However, we cannot predict further events, such as an intense AGN period, that could rapidly accelerate this process.  

Although NGC\,7252 provides a strong proof of concept that two disc galaxies can be transformed into an elliptical on rather short timescales, it cannot be used to study the complete transformational process to quiescence. Detailed observations of similar major merger remnants within a range of several 100 Myr before and a few Gyr after first pericentre passage are still required to constrain physical prescriptions of merger simulations and fully understand the evolution in star formation in galaxy mergers.

\begin{acknowledgements}
We thank the anonymous referee for their comments which improved our original manuscript. We also extend a special thanks to Glenn van de Ven for helpful discussion about the project. JW acknowledges financial support and thanks Max Planck Institute for Astronomy for the great hospitality during his two month summer student internship in 2017, during which the majority of this work was performed. IMN acknowledges funding from the Marie Sk{\l}odowska-Curie Individual Fellowship 702607, and from grant AYA2013-48226-C3-1-P from the Spanish Ministry of Economy and Competitiveness (MINECO). RMcD is the recipient of an Australian Research Council Future Fellowship (project number FT150100333).
\end{acknowledgements}


\bibliographystyle{aa}
\bibliography{references}

\end{document}